\documentclass[a4paper,twocolumn]{esapub} 

\usepackage{epsfig}

\title{Prospects for detectability of classical novae with INTEGRAL}
\author[1,2]{M. Hernanz}
\author[1]{J. G\'omez-Gomar}
\author[1,3]{J. Jos\'e}
\author[4]{A. Coc}
\affil[1]{Institut d'Estudis Espacials de Catalunya, IEEC}
\affil[2]{Instituto de Ciencias del Espacio, CSIC}
\affil[3]{Departament de F\'{\i}sica i Enginyeria Nuclear (UPC)}
\affil[4]{Centre de Spectrom\'etrie
Nucl\'eaire et de Spectrom\'etrie de Masse, IN2P3-CNRS}

\begin{document}

\maketitle


\begin{abstract}
Classical novae are potential gamma-ray emitters, both in lines and in a
continuum. Continuum emission (at energies between 20-30 and 511 keV) and
line emission at 511 keV are related to positron
annihilation and its Comptonization in the expanding shell; $^{18}$F
is the main responsible of positron production. The lines at 478 and 1275
keV have their origin on the decay of the radioactive  nuclei
$^{7}$Be and $^{22}$Na. Updated models of nova explosions have been adopted
for the computation of the gamma-ray emission. New yields of some radioactive
isotopes directly translate into new detectability distances of classical novae
with INTEGRAL.                     
\end{abstract}

\section{Introduction}
\vspace{-0.5cm}
Novae have been claimed as potential gamma-ray emitters since long ago 
(Clayton \& Hoyle 1974; Clayton 1981; Leising \& Clayton, 1987), but they 
have not been detected yet (Harris, Leising \& Share, 1991; Iyudin et al, 
1995, 1999; Leising et al. 1988; Harris et al. 1996, 1999; see also Jean et 
al. these proceedings), probably because the correponding instruments were 
not sensitive enough. 

A detailed analysis of this emission, based on theoretical models for both 
the explosion 
and the production and propagation of gamma-rays, have been presented only 
recently (G\'omez-Gomar et al. 1998; Hernanz et al. 1999a), specially in 
the context of INTEGRAL and COMPTON satellites capabilities
(Hernanz et al, 1997a,b; Hernanz et al. 1999b; Hernanz et al. 2000a). 
In this paper we update our models, taking 
into account recent results from nuclear physics which affect nova yields of 
radioactive isotopes. The detectability distances of novae with the future 
SPI intrument onboard INTEGRAL are updated accordingly.

\section{Nova models}
\vspace{-0.5cm}
The emission of gamma-rays from classical novae is related to the 
radioactive decay of some unstable isotopes, synthesized during the explosion. 
The most relevant ones are: $^{13}$N ($\tau$=862 s), $^{18}$F 
($\tau$=158 min), $^{7}$Be ($\tau$=77 d), $^{22}$Na ($\tau$=3.75 yr), 
together with $^{26}$Al ($\tau$=1.04x10$^6$ yr). 
The amount of radioactive nuclei in the expanding ejecta determines the 
number of available photons and positrons, whereas the physical conditions of 
the ejecta (mainly densities, velocities and chemical composition) determine 
its opacity to positrons and gamma-rays and, 
consequently, the final amount of gamma-ray radiation produced which escapes.
Therefore, complete models of the explosions, yielding the complete physical 
conditions of the ejecta (and, in particular, its content of radioactive 
isotopes) are needed, together with a detailed treatement of gamma-ray 
production and propagation, in order to compute theoretical models of 
gamma-ray emission (see G\'omez-Gomar et al., 1998, for details).

The nova models have been computed with the hydrodynamic code SHIVA (see 
Jos\'e \& Hernanz, 1998, for details), which includes a complete reaction 
network with updated nuclear reaction rates (see Jos\'e, Coc \& Hernanz, 
1999 and Hernanz et al. 1999a). Both CO and ONe novae are considered, 
with initial luminosity 10$^{-2}$L$_\odot$ and accretion rate 
2x10$^{-10}$M$_\odot$/yr. The main properties of the ejecta relevant for 
their gamma-ray emission are shown in table 1; the yields of the short-lived 
isotopes $^{13}$N and $^{18}$F correspond to 1 hour after peak temperature. 
A quick look at the table shows that $^{7}$Be is mainly produced in CO novae, 
whereas $^{22}$Na and $^{26}$Al are mainly synthesized in ONe novae. The 
short lived isotopes $^{13}$N and $^{18}$F are produced by roughly the same 
amounts in both nova types. These results have important consequences in the 
ensuing gamma-ray emission, as shown below.

It is important to notice the origin of the main changes between these 
results and those presented by us some years ago. There have 
been important changes in the nuclear reactions affecting $^{18}$F synthesis. 
$^{18}$F synthesis depends mainly on the rate of the $^{18}$F(p,$\alpha$)
reaction. New rates from Utku et al. (1998) lead to 
reductions of the $^{18}$F yields by factors of around 10 (yields in table 1 
as compared with those in G\'omez-Gomar et al. 1998). The resulting reduction 
in the prompt gamma-ray emission (at energies between 20-30 and 511 keV, see
below) of both CO and ONe 
novae is of the same order. Unfortunately, there is still a large uncertainty,
by a factor of $\sim$300, affecting those rates (see Coc et al. 2000).

For $^{22}$Na and $^{26}$Al, the changes have 
been minor, but also deserve some attention, mainly because there is still 
some uncertainty remaining. The main reaction rates affecting $^{22}$Na 
synthesis are $^{21}$Na(p,$\gamma$) and $^{22}$Na(p,$\gamma$), leading to 
uncertainties in the yields of roughly a factor of 3, although very recently 
the uncertainty of the  $^{21}$Na(p,$\gamma$) reaction has been reduced from 
5 orders of magnitude to 1 (Smirnova \& Coc, 2000). 

\section{Spectra and light curves}
According to its duration and time of appearance,  gamma-ray emission 
produced during nova explosions can be considered as prompt emission and 
long-lasting emission.
The prompt emission is related to positron annihilation, with the 
positrons coming from $^{13}$N and $^{18}$F decays. Prompt emission consists 
of 
a line at 511 keV plus a continuum below this energy, with a cutoff at 20-30 
keV, related to photoelectric absorption. This continuum comes from both 
positronium emission and Comptonization of 511 keV photons (see 
G\'omez-Gomar et al. 1998 and Hernanz et al. 1999a for details). The prompt 
emission is the most intense gamma-ray emission from novae, 
but it has very short duration and appears only a few hours after 
the explosion (defined as time of T$_{\rm peak}$, which occurs before 
the maximum in visual luminosity). That's because of the short decay times 
of $^{13}$N and $^{18}$F. This emission is also strongly dependent on the 
transparency of the model at these very early times. 

The long-lasting emission consists of lines at 478 keV ($^{7}$Be decay) 
and 1275 keV ($^{22}$Na decay). There should also be a line at 1809 keV, 
related to $^{26}$Al 
decay, but the long lifetime of this isotope prevents to observe this line in 
individual objects, since the time interval between two succesive Galactic 
nova explosions is much shorter. Here we concentrate on the emission by 
individual novae and, therefore, we don't take into account $^{26}$Al 
emission. 

In figures 1 and 2 we show the gamma-ray spectra of a CO (1.15 M$_\odot$) 
and two ONe novae (1.15 and 1.25 M$_\odot$) at different epochs 
after the explosion (all these novae belong to the fast or very fast speed 
class, whereas the 0.8 M$_\odot$ one would be slow or moderately fast). 
It can be seen that both nova types 
display a prominent 511 keV line (which disappears very fast, see light 
curves in figure 3). In addition to this line and to the continuum below it, 
CO novae show the 478 keV line during some days (see G\'omez-Gomar et al. 
1998), whereas ONe 
novae show the 1275 keV line, lasting for some months (see figure 4). The 
different features of CO and ONe novae in models considered are strictly 
related to their different nucleosynthesis: 
CO novae produce larger amounts of $^{7}$Be than ONe novae, whereas the 
reverse is true for $^{22}$Na. The effect of the white dwarf mass is shown 
in figure 2 for the ONe novae. In this case, slight differences appear, 
since 
both the content of radioactive elements and the dynamic properties of the 
ejecta are similar. In the CO case, on the contrary, the two chosen models are 
quite different in what concerns the dynamics (the 0.8 M$_\odot$ nova is quite 
opaque, as compared to the 1.15 M$_\odot$ one; see G\'omez-Gomar et al. 1998).
Consequently, the fluxes 
during the prompt emission are smaller. Also, the smaller content of $^{7}$Be 
in the low mass CO case leads to a smaller flux of the 478 keV line (see 
G\'omez-Gomar et al. 1998).

\section{Detectability with INTEGRAL/SPI. Discussion}
The 3$\sigma$ detectability distances of our models with the SPI instrument 
onboard INTEGRAL are shown in table 2. To compute them, we have 
fully taken into account the line profiles and the total integration time.  
The 478 keV line is not very broad (FWHM: 3-7 keV), but the 1275 keV line is 
quite broad (FWHM: 20 keV). This is crucial for the detectability distances, 
since SPI's sensitivity worsens for broad lines. Concerning the 511 keV line, 
its FWHM ranges from 3 to 8 keV, but the short duration of the emission 
prevents to adopt the nominal $10^6$ s observation time. In this case, 10 
hours integration time, starting 5 hours after the explosion, has been 
adopted.

There is an extra problem with the 511 keV line, related to its early 
appearence before optical detection. This leads to detection only possible if 
a nova falls in the FOV of the instrument when it is doing another observation 
(like the Galactic Plane Survey or the Central Radian Deep Exposure, for 
instance). We have also considered alternative ways to detect this intense 
emission, through the SPI's shield (see Jean et al. 1999). It is very 
important to stress that the successful detection of the annihilation emission 
from novae 
strongly relies on the availability of {\it a posteriori} analyses, like those 
we are doing with BATSE data (see Hernanz et al. 2000b) or those done by 
Harris et al. (1999) with the TGRS instrument onboard the WIND satellite. 

In table 2 we also show the detectability distances for the continuum, 
defined to be {\it optimal} for SPI, i.e., between 170 and 470 keV (with 
integration time 10h, like for the 511 keV line).
It is worth mentioning that the flux emitted in the continuum is larger than 
the 
flux in the 511 keV line, but the high spectral resolution of SPI makes its 
detectability harder and, therefore, detectability distances for the continuum 
are not larger than those for the 511 keV line.
However, this intense continuum flux (still larger if the full 
20-511 keV range is included) poses some hope in its future detectability by 
large FOV instruments, doing all-sky surveys in hard X-rays up to some 
hundreds of keV, like the EXIST instrument.

In summary: SPI detectability distances for the 1275 keV line from ONe novae 
are around 1 kpc, during months after the explosion, and detectability 
distances for the 478 keV line from CO novae are around 0.5 kpc, during weeks 
after the explosion. These figures imply that during INTEGRAL's nominal 
lifetime, 2 years, no 
more than one or two novae could be detected in the 1275 keV line (for the 478 
keV line the situation is worse and for the 511 keV one it will depend on some 
yet not fixed procedures). However, since the distance of novae is not well 
determined in many cases (at least before a deep analysis of the 
optical-UV-IR observations 
during some months), conservative assumptions concerning optical 
brightness at discovery-maximum should be adopted, in order not to miss 
{\it the good one}.

\vspace{-0.5cm}
\section*{Acknowledgments}
\vspace{-0.5cm}
Research partially supported by the CICYT-P.N.I.E.
(ESP98-1348), by the DGICYT (PB98-1183-C03-02 and PB98-1183-C03-03) and 
by the AIHF1999-0140


\begin{table*}
\begin{center}
\caption{Radioactivities in novae ejecta}
\vspace{0.3cm}
\renewcommand{\arraystretch}{1.5}
\begin{tabular}{ccccccccc}
\hline
Nova        & M$_{\rm wd}(\rm M_\odot)$ & M$_{\rm ejec}(\rm M_\odot)$ & 
KE (erg/g)  & $^{13}$N (M$_\odot$)      & $^{18}$F (M$_\odot$)           & 
              $^{7}$Be (M$_\odot$)      & $^{22}$Na (M$_\odot$)          &
              $^{26}$Al (M$_\odot$)\\
\hline
CO          & 0.8                       & 6.2x10$^{-5}$                  &
8x10$^{15}$ & 1.5x10$^{-7}$             & 1.8x10$^{-9}$                  &
              6.0x10$^{-11}$            & 7.4x10$^{-11}$                 &
              1.7x10$^{-10}$\\
CO          & 1.15                      & 1.3x10$^{-5}$                  &
4x10$^{16}$ & 2.3x10$^{-8}$             & 2.6x10$^{-9}$                  &
              1.1x10$^{-10}$            & 1.1x10$^{-11}$                 &
              6.1x10$^{-10}$\\
ONe         & 1.15                      & 2.6x10$^{-5}$                  &
3x10$^{16}$ & 2.9x10$^{-8}$             & 5.9x10$^{-9}$                  &
              1.6x10$^{-11}$            & 6.4x10$^{-9}$                  &
              2.1x10$^{-8}$\\
ONe         & 1.25                      & 1.8x10$^{-5}$                  &
4x10$^{16}$ & 3.8x10$^{-8}$             & 4.5x10$^{-9}$                  &
              1.2x10$^{-11}$            & 5.9x10$^{-9}$                  &
              1.1x10$^{-8}$\\
\hline
\end{tabular}
\end{center}
\end{table*}


\begin{table*}
\begin{center}
\caption{SPI 3$\sigma$ detectability distances (in kpc) for lines and 
continuum}
\vspace{0.3cm}
\renewcommand{\arraystretch}{1.5}
\begin{tabular}{cccccc}
\hline
Nova type   & M$_{\rm wd}(\rm M_\odot)$ & 511 keV line & 478 keV line 
            & 1275 keV line             & continuum (170-470) keV\\
\hline
CO          & 0.8                       &   0.7        &   0.4        
            & -                         &   0.4\\
CO          & 1.15                      &   2.4        &   0.5        
            & -                         &   2.0\\
ONe         & 1.15                      &   3.7        &   -          
            & 1.1                       &   3.0\\
ONe         & 1.25                      &   4.3        &   -          
            & 1.1                       &   3.0\\
\hline
\end{tabular}
\end{center}
\end{table*}


\begin{figure}
\begin{center}
\epsfig{file=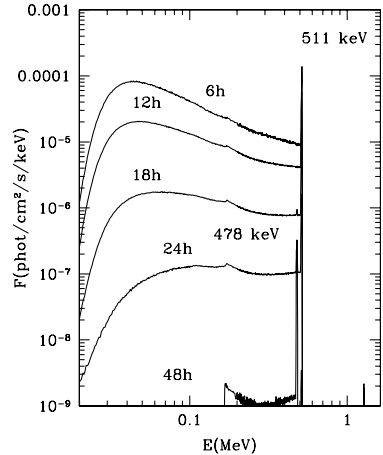,width=8cm}
\vspace{-1.5cm}
\caption{Spectral evolution of a 1.15 M$_\odot$ CO nova}
\end{center}
\end{figure}


\begin{figure}
\begin{center}
\epsfig{file=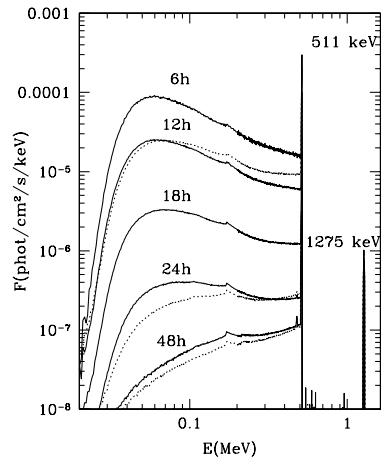,width=8cm}
\vspace{-1.5cm}
\caption{Spectral evolution of 1.15 (solid) and 1.25 M$_\odot$ (dotted) 
ONe novae}
\end{center}
\end{figure}

\newpage


\begin{figure}
\begin{center}
\epsfig{file=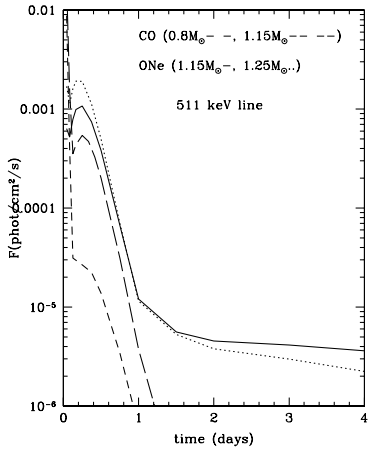,width=8cm}
\vspace{-1.5cm}
\caption{Light curves ot the 511 keV line for CO and ONe novae}
\end{center}
\end{figure}


\begin{figure}
\begin{center}
\epsfig{file=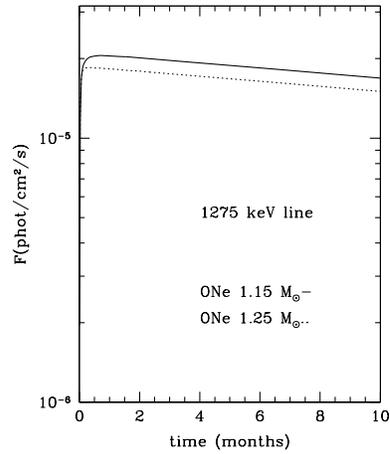,width=8cm}
\vspace{-1.5cm}
\caption{Light curves ot the 1275 keV line for ONe novae}
\end{center}
\end{figure}

\end{document}